\documentclass[sigconf]{acmart}
\usepackage{enumitem}
\usepackage{multirow}

\DeclareGraphicsExtensions{.pdf,.ai,.jpg,.png}
\setkeys{Gin}{pagebox=artbox}

\copyrightyear{2023}
\acmYear{2023}
\setcopyright{licensedothergov}\acmConference[SIGIR-AP '23]{Annual International ACM SIGIR Conference on Research and Development in Information Retrieval in the Asia Pacific Region}{November 26--28, 2023}{Beijing, China}
\acmBooktitle{Annual International ACM SIGIR Conference on Research and Development in Information Retrieval in the Asia Pacific Region (SIGIR-AP '23), November 26--28, 2023, Beijing, China}
\acmPrice{15.00}
\acmDOI{10.1145/3624918.3625322}
\acmISBN{979-8-4007-0408-6/23/11}

\begin{document}
	
\title{Generating Natural Language Queries for More Effective Systematic Review Screening Prioritisation}

\settopmatter{authorsperrow=3}

\author{Shuai Wang}
\affiliation{%
	\institution{The University of Queensland}
  \city{}
  \country{}
}
\email{}

\author{Harrisen Scells}
\affiliation{%
  \institution{Leipzig University}
  \city{}
  \country{}
}
\email{}

\author{Martin Potthast}
\affiliation{%
	\institution{Leipzig University \& ScaDS.AI}
  \city{}
  \country{}
}
\email{}

\author{Bevan Koopman}
\affiliation{%
	\institution{\mbox{\hspace{-4.5pt}CSIRO~\&~The~University~of~Queensland\hspace{-4.5pt}}}
  \city{}
  \country{}
}
\email{}

\author{Guido Zuccon}
\affiliation{%
	\institution{The University of Queensland}
  \city{}
  \country{}
}
\email{}

\newcommand\todo[1]{{\color{red}#1}}

\begin{abstract}
Screening prioritisation in medical systematic reviews aims to rank the set of documents retrieved by complex Boolean queries. Prioritising the most important documents ensures that subsequent review steps can be carried out more efficiently and effectively. The current state of the art uses the final title of the review as a query to rank the documents using BERT-based neural rankers. However, the final title is only formulated at the end of the review process, which makes this approach impractical as it relies on ex post facto information. At the time of screening, only a rough working title is available, with which the BERT-based ranker performs significantly worse than with the final title. In this paper, we explore alternative sources of queries for prioritising screening, such as the Boolean query used to retrieve the documents to be screened and queries generated by instruction-based generative large-scale language models such as ChatGPT and Alpaca. Our best approach is not only viable based on the information available at the time of screening, but also has similar effectiveness to the final title.
\end{abstract}



\maketitle

\section{Introduction}

Systematic reviews are a widely used type of literature review in evidence-based medicine to comprehensively identify, analyse and summarise all available research on a particular topic or question in an unbiased manner~\cite{kitchenham2004procedures}. They provide a rigorous and transparent pathway to medical decision-making tasks and minimise bias and errors that might otherwise result from an ad hoc literature search~\cite{tranfield2003towards}. Systematic reviews are usually conducted according to a protocol of established steps~\cite{kitchenham2004procedures, deeks2022cochrane, higgins2019cochrane}. As part of this process, complex Boolean queries are developed by specialists (e.g. trained experts in search) to obtain a large initial set of (candidate) documents. These candidates are manually screened to obtain a subset for in-depth analysis. One approach to increase the efficiency of relevance assessment in systematic reviews is know as \textit{screening prioritisation}~\cite{karnon2007review, o2015using}. The aim of this method is to rank the candidates so that the subsequent review tasks, such as full text screening, can start earlier and in parallel with the screening, resulting in a more timely and predictable completion. For rapid reviews, where screening is constrained by a limited budget, prioritising screening can help to identify more relevant documents and thus enable a review of higher quality~\cite{tricco2015scoping}. 

Currently, the most advanced screening prioritisation methods use BERT-based neural rankers~\cite{wang2022neural}. They are based on a fundamental but ultimately unjustified assumption that the title of the systematic review has already been conceived and is available as a query for ranking before screening. However, according to the systematic review protocol, the title of the review does not have to be formulated before the search~\cite{deeks2022cochrane, higgins2019cochrane}: as a rule, the title is only determined at the time of writing. Instead, most systematic reviews have only a rough working title, usually just a few keywords of the study~\cite{wang2022little}. Our experiments show that using a working title for screening prioritisation is not competitive to using the final title~\cite{gregory2018introduction}: When applied to the Seed Collection dataset~\cite{wang2022little}, which contains both working and final titles of systematic reviews, the effectiveness of a neural ranker (Section~\ref{sec-neural-ranker}) depends strongly on which of the two title versions is used (Table~\ref{table-oracle-vs-practical-effectivenss}, top rows)~\footnote{Please note, in our ACM published paper, the result of working title in Seed Collection was wrong due to bug in data pre-processing, the is updated here, and the update of the result does not have any influence to the observation and conclusion made from this paper.} . Using the final title largely overestimates the achievable effectiveness.

\begin{table}[t]
\centering
\small
\renewcommand{\tabcolsep}{2.5pt}
\caption{Our contribution at a glance: The post hoc effectiveness of \citeauthor{wang2022neural}'s~\cite{wang2022neural} original approach can be achieved by generating queries from sources available in practice; *~indicates statistical significant differences.}
\vspace{-10pt}
\label{table-oracle-vs-practical-effectivenss}
\begin{tabular}{@{}llccccl@{}}
\toprule
  \bf Source                             & \bf Query            &      \bf MAP      &     \bf LastRel     &     \bf WSS95     &    \bf WSS100     & \bf Ref.              \\
\midrule
  \raisebox{-1.25ex}[0em][0em]{post hoc} & final review title   &       0.295       &       634.975       &  \textbf{0.609}   &  \textbf{0.597}   & \cite{wang2022neural} \\
                                         & best generated query &  \textbf{0.310}   &  \textbf{620.025}   &       0.589       &       0.569       & \bf ours              \\
\midrule
  \raisebox{-1.25ex}[0em][0em]{practice} & working title       & 0.171*\kern-0.3em & 801.050 *\kern-0.3em & 0.465*\kern-0.3em & 0.450*\kern-0.3em & \cite{wang2022neural} \\
                                         & generated\,queries   &       0.249       &       714.500       & 0.541*\kern-0.3em & 0.521*\kern-0.3em & \bf ours              \\
\bottomrule
\end{tabular}
\vspace{-10pt}
\end{table}

In this paper, we investigate how screening prioritisation can be done based on the information available at the time of screening. Since it can take up to several weeks to work out the Boolean queries for the prior retrieval of candidate documents~\cite{saleh2014grey}, we propose effective ways to exploit this valuable source of information that has so far been neglected by the state of the art~\cite{wang2022neural, lee2018seed, wang2022seed, carvallo2020automatic, karimi2010boolean, lee2017study}. However, using a Boolean query to produce a ranked list is not straightforward. A Boolean query is complex, structured, and detailed; it is very different from the queries that are common in ad hoc retrieval~\cite{bramer2018systematic}. BERT-based methods for ranking may perform poorly on these queries. Therefore, we investigate the use of two instruction-based models, namely OpenAI's ChatGPT~\cite{gozalo2023chatgpt} and Stanford's Alpaca~\cite{alpaca}, to generate natural language queries from Boolean queries. These generated natural language queries are in turn used as input for our neural-ranker-based screening prioritisation methods. The bottom rows of Table~\ref{table-oracle-vs-practical-effectivenss} show that the most powerful variants of our method are able to generate queries that compete with the use of the final title.%
\footnote{Code: \url{https://github.com/ielab/SIGIR-AP-2023-Bolean2Natural4SR}}
To guide our investigation, we have developed five research questions:
\begin{itemize}
\setlength{\itemindent}{0em}
\item[\textbf{RQ1}]
How effective is screening prioritisation with Boolean queries compared to natural language queries generated from them?
\item[\textbf{RQ2}]
How do different generation models affect the effectiveness of natural language queries generated from Boolean queries?
\item[\textbf{RQ3}]
What impact do ranking methods have on the effectiveness of natural language queries derived from Boolean queries?
\item[\textbf{RQ4}]
Does generating multiple natural language queries from a single Boolean query improve effectiveness?
\item[\textbf{RQ5}]
How effective is screening prioritisation with natural language queries derived from Boolean queries compared to using the working titles of systematic reviews?
\end{itemize}

\section{Related Work}

\begin{figure*}
\includegraphics{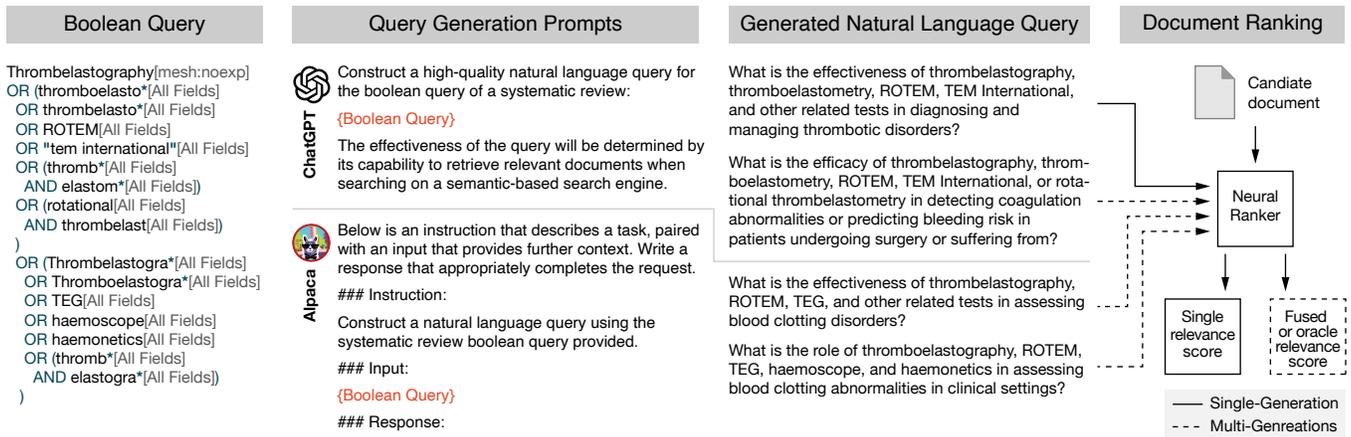}
\caption{Illustration and examples of our screening prioritisation approach: Given a Boolean query, an instruction-based LLM is prompted to generate one or more natural language queries. Then, given a generated query and a candidate document, a neural ranker is used to predict one or more relevance scores for the document. In the latter case, the scores are fused by addition. As a baseline for our experiments, the score that maximises effectiveness is selected by an oracle.}
\label{fig:architecture}
\end{figure*}

In this section, we review the literature on screening prioritisation for systematic reviews and instruction-based large language models.

\subsection{Systematic Review Screening Prioritisation}

Screening prioritisation has received considerable attention in tech\-nology-assisted systematic review generation. Various aspects were investigated, including the use of different input data sources~\cite{lee2018seed, wang2022seed, scells2020clf}, different algorithms or models for ranking purposes~\cite{lee2017study, kanoulas2018clef, wang2022neural, wang2022little, minas2018aristotle, chen2017ecnu, norman122017limsi, scells2017ltr, alharbi2017ranking, alharbi2018retrieving}, and active learning techniques that improve the efficiency of screening prioritisation through a human-in-the-loop approach~\cite{carvallo2020automatic, di2018interactive, di2017interactive, minas2018aristotle, anagnostou2017combining, yu2017data, singh2017iiit, norman2018limsi, hollmann2017ranking, yang2022goldilocks}.

\textit{Boolean-driven screening prioritisation} uses a Boolean query to rank candidate documents directly. While few studies have examined using Boolean queries alone, most used them in conjunction with the final review title~\cite{alharbi2017ranking, alharbi2018retrieving, alharbi2019ranking}. Typically, keywords are extracted from the Boolean query and the review title to formulate a (bag of words) query, and then a lexical scoring function determines the relevance of a document to the query. However, these methods are impractical since the final title is not available at the time of screening. The coordination level fusion~(CLF) approach proposed by~\citet{scells2020clf} is the only existing method that examines screening prioritisation using Boolean queries only. It uses rank fusion to rank the documents retrieved by each clause of a Boolean query. 

\textit{Neural ranker-based screening prioritisation} methods rely on pre-trained models such as BERT and have achieved much higher effectiveness than traditional lexical rankers, on par with active learning methods that use relevance signals from the screened documents~\cite{wang2022neural}. Despite the improvements they have brought to screening prioritisation, there are still challenges in using them: For instance, the input token length limitation imposed by most BERT-based models~\cite{devlin2018bert} is a critical limitation. It does not allow the model to process longer text inputs, such as the full text of candidate documents, extensive Boolean queries, or seed studies as a source of information. Previous approaches using neural rankers for screening prioritisation has focused only on using the review title as a query. We show that their effectiveness does not generalise when working titles are used instead (see Table~\ref{table-oracle-vs-practical-effectivenss}).

\subsection{Instruction-based Large Language Models}

Recent advances in instruction-based large language models~(LLMs), such as ChatGPT, have shown that they are able to accurately follow user instructions to complete tasks~\cite{gozalo2023chatgpt, guo2023close,wang2023can, sallam2023chatgpt}. These models typically contain tens of billions of parameters and are trained on diverse and extensive textual data so that they are able to generate relevant and coherent answers for a wide range of topics~\cite{gozalo2023chatgpt}. Several studies have evaluated the effectiveness of ChatGPT on various tasks, often observing an increase in effectiveness compared to previous approaches, e.g., in question answering~\cite{tan2023evaluation, omar2023chatgpt} or ranking~\cite{sun2023chatgpt, ji2023exploring}. As part of a systematic review literature search, the use of ChatGPT to generate Boolean queries for systematic reviews has been investigated~\citet{wang2023can}. The results of this study showed that ChatGPT generates effective queries with appropriate prompting. In this paper we use ChatGPT and Alpaca~\cite{alpaca}. Alpaca was fine-tuned based on an LLM developed by Meta with seven billion parameters, known as LLaMa~\cite{touvron2023llama, alpaca}. Alpaca has been fine-tuned using 52K~instruction--output pairs generated from ChatGPT via a self-instruction approach~\citet{wang2022self}, showing similar capabilities to ChatGPT in preliminary human evaluations~\cite{alpaca}.

For ranking tasks, instruction-based language models are integrated with ranking models to achieve more effective results~\cite{wang2023query2doc, gao2022precise, jiang2023active, yasunaga2023retrieval}. Specifically, there are two common ways to combine these models: \textit{retrieval-then-generation} and \textit{generation-then-retrieval}. In the \textit{retrieval-then-generation} approach, the ranking model first retrieves a set of relevant results based on the user's query. Then, the instruction-based language model generates a response based on the retrieved documents. This method relies primarily on the ranking model's ability to understand the user and extract corresponding information from their query, while the LLM is used to summarise the retrieved evidence to provide the user with a credible and comprehensive answer~\cite{jiang2023active, yasunaga2023retrieval}. In the \textit{generate-then-retrieval} approach, the instruction-based LLM is first used to generate a response based on the user's query, and this response is then processed by a ranking model as a new query to retrieve documents that provide evidence for its statements~\cite{wang2023query2doc, gao2022precise}.

\section{Methodology}

\enlargethispage{\baselineskip}
Figure~\ref{fig:architecture} gives an overview of our approach for screening prioritisation. One or more natural language queries are generated from a given Boolean query. Then the candidate documents to be screened are ranked based on the generated queries.

\subsection{Query Generation}
\label{sec:bool_query_conversion}

Our task is to generate a natural language query from a Boolean query of a systematic review that best describes its information need. We evaluate two LLMs for this task: ChatGPT~\cite{gozalo2023chatgpt}%
\footnote{We used the OpenAI's GPT-3.5-turbo API with a maximum of 4,097~tokens.}
and Alpaca~\cite{alpaca}.%
\footnote{The model was fine-tuned using the original setup of Stanford's Alpaca.}
Figure~\ref{fig:architecture} shows an example consisting of a Boolean query, carefully optimized prompts for each of the two models, and four alternative generated natural language queries. Preliminary studies have shown that Alpaca has problems with zero-shot generation for virtually all Boolean queries, often returning the original Boolean query itself. We addressed this problem by fine-tuning Alpaca using pairs of Boolean queries and natural language queries generated by ChatGPT as training examples.

To adjust the ``creativity'' of an LLM, they often introduce a degree of stochasticity controlled by the so-called temperature parameter~$t$, where~$0 \leq t \leq 1$~\cite{levy-etal-2021-investigating}. Setting $t = 0$ causes the model to generate the same response over multiple inferences for a given prompt, whereas $t = 1$ causes the model's response to be randomly different each time. In other words, the lower the temperature value, the more deterministic a model's response. In our experiments, we investigate how the creativity of a model affects the effectiveness of screening prioritisation~(RQ4). Therefore, we compare the generation of a single natural language query (\textit{Single-Generation}) to generating multiple natural language queries (\textit{Multi-Generations}) by adjusting the temperature accordingly.%
\footnote{As Hugging Face does not allow $t=0$, we use $t=0.0001$ instead.}

\subsection{Document Ranking}

To rank the documents, we follow the state-of-the-art screening prioritisation method developed by~\citet{wang2022neural}. Here, a cross-encoder-based neural ranker is used to calculate the relevance score of a query--document pair. Specifically, the query and the document are first concatenated with a $[SEP]$~token, and then fed into a cross-encoder model that calculates the relevance score for the concatenated pair. The relevance score is then represented by the special classification token~$[CLS]$ in the output of the model~\cite{clark2019does}. In the proposed pipeline, the query is obtained from a Boolean query as described above. However, in our experiments, we also explore the alternative of using the original Boolean query as input to the cross-encoder to address~RQ1, and the alternative of using the working title of the review to address~RQ5.  

For fine-tuning, we first use a pre-trained BioBERT model, which has been shown to be effective in screening prioritisation when the title of the review is used as query~\cite{wang2022neural}. Then, for each topic in the training set, we extract all relevant documents~$D^+$ and a number of non-relevant documents~$D^-$. For each pair of relevant document~$(d^+,d^-) \in D^+ \times D^-$, we create training triples $\langle \mathrm{query}, d^+, d^- \rangle$ and then fine-tune the model using localised contrastive loss as proposed by~\citet{gao2021rethink}.

To investigate RQ4, we generate multiple queries per Boolean query in the Multi-Generation setup, calculate a relevance score for each pair of query and candidate document, and then apply two strategies to derive a single relevance score from them for a candidate document: \textit{Fusion} and \textit{Oracle} selection. In the \textit{Fusion} strategy, the relevance scores of all natural language queries on the same topic that refer to a candidate document are summed to calculate the final relevance score of the document with respect to the topic of the systematic review. For the \textit{Oracle} strategy, we first evaluate the ranked lists from different natural language queries, after which the best-performing ranked list, as measured by the mean average precision (MAP), is selected. This strategy serves as an upper bound baseline.

To investigate the effectiveness of combining the results of a generated natural language query with those of the original Boolean query, we also evaluate a setup that includes a fusion of their ranking results. For this purpose, we use the COMBSUM fusion technique to fuse the two ranked lists~\cite{fox1994combination}: the relevance score of a document in the fused ranked list is the sum of the individual scores of the document in the two lists to be fused.

\section{Experimental Setup}

In this section, we outline the datasets we use, the methods we apply, and how we evaluate them.

\subsection{Dataset}

We use two collections in our experiments. The \textit{CLEF TAR Collection} comprises three datasets from~2017, 2018, and~2019. In~2017, the dataset includes 50~systematic review topics divided into 20~for training and 30~for testing~\cite{kanoulas2017clef}. In~2018 the dataset is expanded including all 50~systematic review topics from~2017 as a training set and adding 30~new topics for testing~\cite{kanoulas2018clef}. The~2017 and~2018 datasets focus on Diagnostic Test Accuracy~(DTA) systematic reviews. The 2019~dataset is divided into four categories of systematic reviews: the DTA~category, which builds upon the 2018~dataset and uses it as the training set with eight new topics for testing; the Intervention category, containing 20~training topics and 20~testing topics; the Prognosis Review category and the Qualitative Review category, each featuring one topic~\cite{kanoulas2019clef}. In our experiments, we treat DTA and Intervention topics as two sub-collections, denoted as CLEF-2019-DTA, and CLEF-2019-Intervention. Each topic in the CLEF TAR Collection provides the review title, the Boolean query used for document retrieval, the documents retrieved as a result of the Boolean query, and the relevance labels for the documents at both abstract and full-text levels~\cite{kanoulas2017clef, kanoulas2018clef, kanoulas2019clef}.

The \textit{Seed Collection} contains 40~systematic review topics without training or testing portions~\cite{wang2022little}. The dataset also contains the review title, the Boolean query used during retrieval, documents retrieved, and relevance labels. However, unlike the CLEF TAR Collection, where abstract-level relevance and full-text level relevance are all included, the Seed Collection only contains full-text level relevance judgements directly extracted from published reviews. One major difference between the Seed Collection and the CLEF TAR Collection is that the dataset also includes more details of the review. For example, it includes a temporary working title for each review, named `search name' in the collection, and a set of seed studies used for Boolean query creation~\cite{wang2022little}.

Unlike previous studies that used only the training portions specified in each dataset, we re-split our training data to include distinct topics from all other datasets (CLEF TAR Collection and Seed Collection) that are not included in the test portion of the respective dataset. We chose this strategy due to the uneven allocation of training data across the datasets we use. For instance, the Seed Collection dataset contains no training topics, whereas CLEF-2017 comprises 20~training topics, and CLEF-2019-DTA holds 80~topics. By incorporating training data from a range of sources, we aim to establish a more balanced and comprehensive training environment for our fine-tuned models.

\subsection{Baseline Methods}

In our experiment, we employ BM25 and the Query Likelihood Model (QLM) as baseline ranking models~\cite{robertson1994some, jay1998qlm}. For query preprocessing, we begin by removing all field types in the query, leaving us with only the query terms. We then apply the matching algorithm to the candidate document, calculating a relevance score between the query and the document.

Similar to previous studies comparing neural rankers with traditional term-matching rankers, we utilise specific tools to implement our baseline models. For~BM25, we employ the Gensim toolkit, an open-source library that offers robust implementations for a variety of information retrieval tasks~\cite{vrehuuvrek2011gensim}. For the~QLM, we apply Jelinek-Mercer~(JM) smoothing, a popular technique for query likelihood estimation~\cite{jay1998qlm}.

In addition to the traditional ranking models, we also benchmark our models against the best-performing methods from participant runs in each CLEF-TAR dataset. It is important to note that certain participant runs have utilised relevance signals from relevance assessments to actively re-rank the remaining documents. We have excluded these runs from our baseline comparison, as they do not align with the scope of our screening prioritisation task, making the comparison unfair. The following participant runs have been selected as baselines for our study: CLEF-2017: \textit{sheffield.run4}~\cite{alharbi2017ranking}; CLEF-2018: \textit{shef-general}~\cite{alharbi2018retrieving}; CLEF-2019-dta: \textit{Sheffield/DTA/DTA\_sheffield-Odds\_Ratio}~\cite{alharbi2019ranking}; CLEF-2019-intervention: \textit{Sheffield/DTA/DTA\_shef\-fieldLog\_Likelihood}~\cite{alharbi2019ranking}.

Lastly, for the CLEF-2017 and~2018 datasets, we compare our method with the CLF~approach proposed by~\citet{scells2020clf}. The CLF~approach stands out as the only existing methodology that has explored the application of Boolean queries for systematic review screening prioritisation.

\subsection{Model Fine-tuning}

In our experiments, we focus on fine-tuning two models: the Alpaca model for query generation and BioBERT for document ranking.

\subsubsection{Fine-tuning the Alpaca Model}

To fine-tune the Alpaca model, our first step involves using Single-Generation to convert the Boo\-lean query into a natural language query for the training portion of each dataset. We use ChatGPT for this conversion task, and consider its output as the gold standard for the Alpaca model to learn from. Following this, we use the prompts shown in the second column of Figure~\ref{fig:architecture} to further fine-tune the Alpaca model to generate a natural language query using the Boolean query of a topic. As Boolean queries for systematic reviews are complex and require many tokens, we opted to simplify the prompt used for Boolean query conversion in ChatGPT. To ensure minimal loss of information from the Boolean query, we increased the input token limit of the Alpaca model from 512 to 768.%
\footnote{768 is the maximum token limit for three 80GB Nvidia A100 GPUs (batch size=1).}
Our fine-tuning process for each Alpaca model continues over three epochs, with batch size and gradient accumulation steps of one each, using three Nvidia 80GB A100 GPUs. For the remaining parameters, we adhered to those used in the original Alpaca work~\cite{alpaca}. During inference, we use the same prompt as fine-tuned to convert the Boolean query to a natural language query in each test dataset.

\subsubsection{Fine-tuning the neural ranker}
\label{sec-neural-ranker}

In our experiments, we chose BioBERT as our pre-trained language model to fine-tune for ranking~\cite{luo2022biogpt}. Previous research has demonstrated that BioBERT shows higher effectiveness in the task of title-driven screening prioritisation~\cite{wang2022neural}. Same as the previous work, we utilise the Reranker toolkit~\cite{gao2021rethink} to fine-tune our model across 100 epochs. The key distinction in fine-tuning and inference pipeline lies in the maximum query length set for all models that utilise Boolean or natural language queries. Instead of the query limit of 64 that was set in previous work for the review title, we extend this to 256 to accommodate the naturally longer input derived from Boolean queries. This adjustment ensures that our models are capable of processing and learning from the full complexity of these queries, potentially enhancing their performance and the accuracy of their outputs.

\subsection{Evaluation}

For the CLEF-TAR Collection, we rely on abstract-level relevance to ensure a fair comparison with the submitted runs. However, for the Seed Collection where no abstract-level labels were provided, we utilise full-text level relevance signals.

To demonstrate the effectiveness of document ranking on screening prioritisation, we compute various evaluation metrics as established in at CLEF TAR. These metrics include Average Precision~(AP), the rank of the last relevant document (Last\_rel), Recall at several percentage cutoffs~(1\%, 5\%, 10\%, and~20\%), and Work Saved over Sampling~(WSS) at~95\% and~100\%. In accordance with the CLEF TAR tasks, we have used the same metrics for evaluating our work and used the tar-2018 evaluation script to evaluate our results~\cite{kanoulas2018clef}.

\section{Main Results}
\label{sec:result}

In this section, we outline and interpret the results from our experiments. Specifically, we delve into the results derived from \textit{Single-Generation} in Section~\ref{subsec:single}, while Section~\ref{subsec:multi} is devoted to examining \textit{Multi-Generations}. Lastly, we perform an ablation study in Section~\ref{subsec:ablation} to further investigate the effectiveness of our method under various experimental configurations.

\begin{table*}
\centering
\small
\caption{Evaluation results for comparing methods for Boolean-driven screening prioritisation by generating natural language queries. We use natural language queries generated by ChatGPT and Alpaca, and the fusions of Boolean/ChatGPT and Boolean/Alpaca. Statistical significant differences (Student's two-tailed paired t-test with Bonferroni correction, $p < 0.05$) between using the Boolean query with the BioBERT ranker, and other approaches are indicated by~$*$.}
\label{table:neural_vs_term_based}
\renewcommand{\tabcolsep}{9pt}
\renewcommand{\arraystretch}{0.99}
\begin{tabular}{@{}ll@{\quad}lllllllll@{}}
\toprule

\bf Dataset&\bf Query&\bf Ranker&\bf MAP&\bf Last\_Rel& \multicolumn{4}{@{}c@{}}{\bf Recall@$x$}&\kern-0.4em\bf WSS95&\kern-0.7em\bf WSS100 \\
\cmidrule(l@{\tabcolsep}r@{\tabcolsep}){6-9}
&&&&&\bf $x=1\%$&\bf $x=5\%$&\bf $x=10\%$&\bf $x=20\%$&& \\
\midrule

\multirow{9}{*}{CLEF-2017} & Boolean& BM25 & 0.114* & 3242.733* & 0.083* & 0.215* & 0.324* & 0.491* & 0.252* & 0.188* \\
& Boolean& QLM & 0.122* & 3223.400* & 0.073* & 0.209* & 0.325* & 0.476* & 0.243* & 0.195* \\
&Boolean & CLF &  0.217 & 3028.033* & 0.149 & 0.341* & 0.473* & 0.671* & 0.442* & 0.327* \\
&\multicolumn{2}{l}{Best Participation Run} & 0.218 & 2382.467* & 0.131 & 0.332* & 0.499* & 0.688* & 0.488* & 0.395* \\
\cmidrule(l@{\tabcolsep}){2-11}
& Boolean&BioBERT & 0.278 &1790.867 & 0.166 & 0.488 & 0.656& 0.812 & 0.600& 0.536\\
& ChatGPT &BioBERT& 0.293 & 1991.167 & 0.150 & 0.476 & 0.643 & 0.801 & 0.590 & 0.501 \\
& Boolean/ChatGPT&BioBERT & \textbf{0.300*} & 1843.133 & 0.170 & \textbf{0.499} & \textbf{0.664} & 0.823 & 0.610 & 0.532 \\
& Alpaca&BioBERT & 0.284 & 1866.000 & 0.165 & 0.435 & 0.607 & 0.789 & 0.591 & 0.502 \\
& Boolean/Alpaca&BioBERT & 0.295 & \textbf{1759.233} & \textbf{0.171} & 0.483 & 0.663 & \textbf{0.827*} & \textbf{0.615} & \textbf{0.539} \\

\midrule

\multirow{9}{*}{CLEF-2018}  & Boolean& BM25 & 0.154* & 6033.067* & 0.082* & 0.242* & 0.391* & 0.563* & 0.361* & 0.264* \\
& Boolean& QLM & 0.157* & 6097.133* & 0.080* & 0.252* & 0.380* & 0.557* & 0.384* & 0.251* \\
&Boolean & CLF & 0.272* & 5743.267* & 0.152 & 0.393* & 0.546* & 0.729* & 0.552* & 0.411* \\
&\multicolumn{2}{l}{Best Participation Run}& 0.258* & 5519.200 & 0.129* & 0.383* & 0.545* & 0.729* & 0.552* & 0.431 \\
\cmidrule(l@{\tabcolsep}){2-11}
& Boolean&BioBERT & 0.353 & 4830.933 & 0.202 & 0.517 & 0.681 & 0.845 & 0.656 & 0.503 \\
& ChatGPT &BioBERT& 0.381 & \textbf{4508.933} & 0.247* & \textbf{0.555*} & \textbf{0.713*} & \textbf{0.865} & \textbf{0.692*} & 0.528 \\
& Boolean/ChatGPT&BioBERT & \textbf{0.386*} & 4603.767* & \textbf{0.247*} & 0.551* & 0.705* & 0.859* & 0.685* & \textbf{0.537*} \\
& Alpaca&BioBERT & 0.333 & 4957.233 & 0.191 & 0.493 & 0.662 & 0.827 & 0.640 & 0.485 \\
& Boolean/Alpaca &BioBERT& 0.365 & 4628.233 & 0.220 & 0.525 & 0.688 & 0.849 & 0.668 & 0.523 \\
\midrule

\multirow{8}{*}{CLEF-2019-DTA}& Boolean & BM25 & 0.125* & 2766.875* & 0.068 & 0.163* & 0.303* & 0.463* & 0.299* & 0.163* \\
& Boolean& QLM & 0.121* & 2614.750* & 0.042 & 0.185* & 0.278* & 0.432* & 0.271* & 0.180* \\
&\multicolumn{2}{l}{Best Participation Run}& 0.248 & 2183.500 & 0.168 & 0.439 & 0.594 & 0.742 & 0.490* & 0.347* \\
\cmidrule(l@{\tabcolsep}){2-11}
& Boolean &BioBERT & \textbf{0.272} & 1146.000 & 0.174 & 0.419 & 0.565 & 0.751 & 0.651 & 0.528 \\
& ChatGPT &BioBERT & 0.247 & 1173.250 & 0.183 & \textbf{0.454} & \textbf{0.594} & \textbf{0.757} & 0.660 & 0.528 \\
& Boolean/ChatGPT &BioBERT & 0.268 & \textbf{1134.375} & \textbf{0.183} & 0.446 & 0.584 & 0.755 & \textbf{0.665} & \textbf{0.545} \\
& Alpaca &BioBERT & 0.241 & 1217.875 & 0.170 & 0.483 & 0.622 & 0.784 & 0.666 & 0.520 \\
& Boolean/Alpaca &BioBERT & 0.251 & 1146.125 & 0.173 & 0.458 & 0.592 & 0.783 & 0.659 & 0.537 \\
\midrule

\multirow{8}{*}{\parbox{1.5cm}{\raggedright CLEF-2019-Intervention}} & Boolean& BM25 & 0.154* & 1479.450* & 0.070* & 0.181* & 0.264* & 0.417* & 0.289* & 0.264* \\
& Boolean& QLM & 0.148* & 1473.850* & 0.041* & 0.201* & 0.287* & 0.450* & 0.295* & 0.252* \\
&\multicolumn{2}{l}{Best Participation Run} & 0.293 & 1132.000 & 0.165 & 0.419 & 0.542 & 0.722 & 0.458 & 0.381* \\
\cmidrule(l@{\tabcolsep}){2-11}
& Boolean&BioBERT  & 0.389 & 1064.300 & 0.195 & 0.458 & 0.619 & 0.733 & 0.557 & 0.499 \\
& ChatGPT&BioBERT  & 0.433* & 993.300 & 0.217 & 0.487 & \textbf{0.654} & 0.788* & 0.573 & 0.503 \\
& Boolean/ChatGPT &BioBERT & \textbf{0.446*} & \textbf{975.750} & \textbf{0.233} & \textbf{0.508*} & 0.651* & \textbf{0.789*} & \textbf{0.578} & \textbf{0.529} \\

& Alpaca &BioBERT & 0.317 & 1087.300 & 0.120 & 0.337* & 0.510* & 0.656 & 0.491 & 0.448 \\		
& Boolean/Alpaca &BioBERT& 0.377 & 1024.700 & 0.169 & 0.460 & 0.616 & 0.730 & 0.551 & 0.504 \\
\midrule

\multirow{7}{*}{Seed Collection} & Boolean& BM25 & 0.087* & 990.100* & 0.034* & 0.140* & 0.249* & 0.412* & 0.252* & 0.253* \\
& Boolean& QLM & 0.085* & 986.475* & 0.018* & 0.141* & 0.212* & 0.397* & 0.254* & 0.260* \\
&Working Title&BioBERT& 0.171* & 801.050* & 0.090* & 0.275* & 0.350* & 0.562* & 0.465* & 0.450*\\
\cmidrule(l@{\tabcolsep}){2-11}
&  Boolean &BioBERT & 0.199 & 785.900 & 0.085 & 0.248 & 0.412 & 0.600 & 0.481 & 0.467 \\
& ChatGPT &BioBERT & 0.217 & \textbf{727.150} & 0.082 & \textbf{0.330*} & \textbf{0.482*} & 0.670* & 0.530* & 0.505 \\
& Boolean/ChatGPT &BioBERT & 0.219 & 744.775* & 0.078 & 0.279 & 0.468* & \textbf{0.677*} & 0.525* & \textbf{0.506*} \\
& Alpaca &BioBERT & 0.221 & 780.600 & 0.083 & 0.314* & 0.473 & 0.655 & 0.529* & 0.500 \\
& Boolean/Alpaca &BioBERT& \textbf{0.230*} & 765.925 & \textbf{0.098} & 0.268 & 0.466* & 0.661* & \textbf{0.531*} & 0.505* \\
\bottomrule

\end{tabular}
\end{table*}

\subsection{Effectiveness of Single-Generation}
\label{subsec:single}

To understand the effectiveness of \textit{Single-Generation}, Table~\ref{table:neural_vs_term_based}~\footnote{Please note, in our ACM published paper, the result of working title in Seed Collection was wrong due to bug in data pre-processing, the is updated here, and the update of the result does not have any influence to the observation and conclusion made from this paper.} compares the ranking effectiveness of the generated query to the original Boolean query, our baseline methods, and title-driven methods (where the working title is used to rank candidate documents). We also evaluate the differences in effectiveness of screening prioritisation between queries generated by various generation models. 

\subsubsection{Boolean vs. Generated Query}

First, we explore the overall effectiveness of neural-ranker-based screening prioritisation using the original Boolean queries versus generated natural language queries. The results suggest that transforming a Boolean query into a natural language query enhances the effectiveness of systematic review screening prioritisation. The only exception to this improvement is seen in CLEF-2019-DTA,%
\footnote{Note that the CLEF-2019-DTA dataset is notably smaller, containing only eight topics (30 topics for other datasets on average), which may make it vulnerable to outliers.}
when using MAP. When evaluating using the recall and WSS measures, generating a natural language query for screening prioritisation achieves higher effectiveness on ranking non-relevant documents at the bottom of the ranking, as denoted by a higher value of Recall@{5\%, 10\%, 20\%}, and WSS{95, 100}; but generally lower effectiveness of Recall@1\%. 

We also find that fusing the ranking results of the generated query with those from the Boolean query further improves effectiveness. Fusion leads to a significantly better ranking than when using the Boolean query alone, particularly for CLEF-2017, 2018, and 2019-Intervention. This finding points to the potential benefits of using a fusion of converted natural language and Boolean queries to improve the ranking of systematic review screening.

\begin{figure*}
\includegraphics[width=\textwidth]{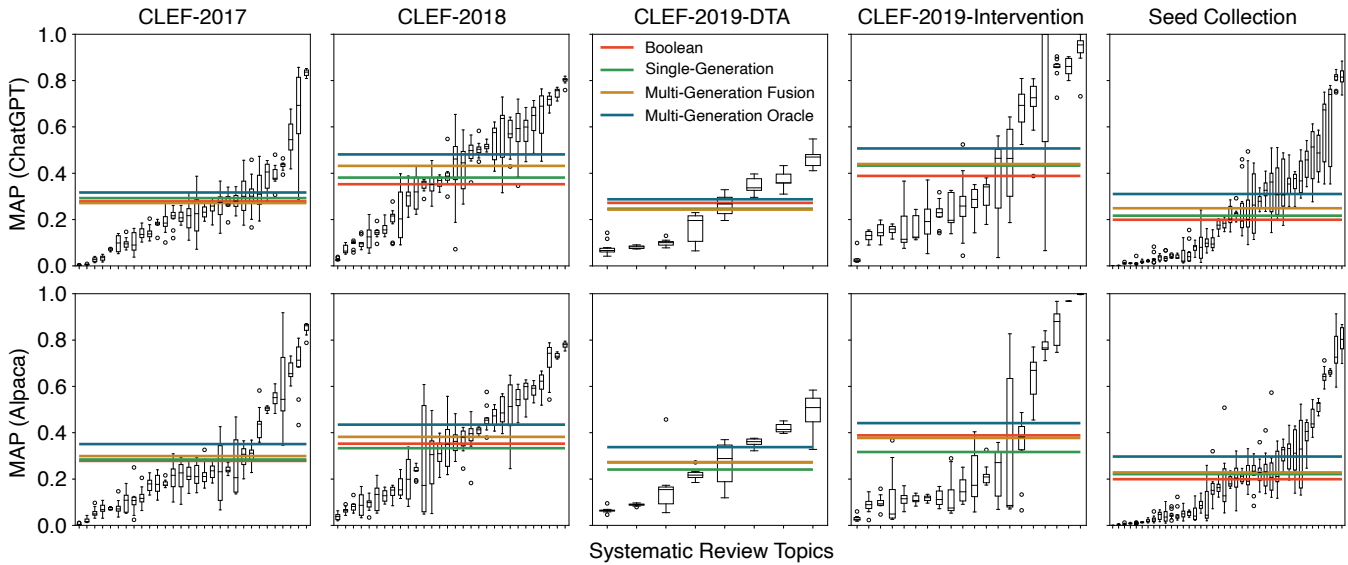}
\caption{Topic-by-topic variability graph for the effectiveness of the Multi-Generations setup, using a single generated natural language query to rank documents. The coloured horizontal lines indicate the average effectiveness of different methods (Boolean, Single-Generation, Multi-Generation Fusion, and Multi-Generation Oracle).}
\label{fig:variability}
\end{figure*}

\subsubsection{Neural vs. Baselines}

When comparing neural-based rankers with lexical methods, we observe that the results from neural-based rankers significantly outperform those from BM25 or QLM when the Boolean query alone is used to rank candidate documents. There are only two exceptions: one in CLEF-2018 when comparing the rank of the last relevant document (Last\_rel), and the other in CLEF-2019-DTA when comparing Recall@1\%. Even in these instances, although higher effectiveness was achieved, it did not reach statistical significance. These results highlight the substantial potential of neural rankers for boolean-driven screening prioritisation.

When we compare our approach to the CLF method, which also only uses Boolean queries for screening prioritisation, we find that our methods exhibit statistically higher effectiveness (except for WSS95 at CLEF-2017 and WSS100 at CLEF-2018). However, the margin is narrower than for methods like BM25 or QLM. Similarly, our methods consistently achieved higher effectiveness over the best participation runs from CLEF. However, similar to the previous comparison, the margin is narrower than when compared to BM25 or QLM, and the difference is not statistically significant in terms of MAP, except for the topics in the CLEF-2018 dataset. While our approach only used the Boolean query for screening prioritisation, the top CLEF entries typically utilised additional input sources, such as the final title of the review, which again shows that the neural method could be beneficial to screening prioritisation. 

\subsubsection{Comparison with using the Working Title}

We are able to compare Boolean-driven screening prioritisation with working-title-driven screening prioritisation exclusively through the Seed Collection, as it is the only collection that provides systematic review working titles. To make this comparison, we trained an additional BioBERT ranker that uses the title from the CLEF dataset to prioritise relevant documents, with the same fine-tuning parameters as previous work~\cite{wang2022neural}. Our findings suggest that although past studies have demonstrated substantial improvements in screening effectiveness when using the final review title as a query, this improvement does not extend to working titles. Remarkably, using working titles results in significantly lower effectiveness than Boolean-driven screening prioritisation methods and even underperforms when compared to basic term-matching methods.

\subsubsection{ChatGPT vs. Alpaca}

Comparing the effectiveness of natural language queries generated by ChatGPT and Alpaca, our results indicate that ChatGPT often outperforms Alpaca in terms of MAP, with the sole exception being the Seed Collection. A significant discrepancy is also noted in CLEF-2019-Intervention, where the natural language query generated by the Alpaca model considerably underperforms compared to both the original Boolean query and the query generated by ChatGPT. This effectiveness drop could be attributed to the difference in systematic review types in the dataset to other datasets (intervention versus DTA). Therefore, the Alpaca model, which has learned to generate queries based on DTA, may not be as effective for intervention topics.

Similar to the ChatGPT-generated queries, those from Alpaca also achieve higher effectiveness when fused with the results from the Boolean query. This approach leads to higher effectiveness across all datasets compared to using only the Boolean query. Notably, in the Seed Collection, the fusion of derived and Boolean queries achieves significantly higher effectiveness compared to the Boolean query alone. This provides further evidence that the Alpaca model can be trained to generate high-quality natural language queries from Boolean queries, equalling the effectiveness of ChatGPT. While Alpaca provides a degree of transparency in its process, unlike ChatGPT, making this comparison even more compelling.

\subsection{Variability and Impact of Multi-Generations}
\label{subsec:multi}

Figure~\ref{fig:variability} shows the results of multiple generations from both the ChatGPT and Alpaca models. Here, both the average effectiveness and the per-topic effectiveness are measured by the MAP metrics. Our findings reveal high variation in effectiveness when using converted queries from both models, with intervention queries appearing as the most unstable topics. We note that extreme variability in effectiveness has also been observed in previous work for both user-edited and system-generated queries, across a range of search domains~\cite{moffat2015pooled,zuccon2016query,palotti2016assessors,liu2019comparative,wang2022automated,scells2019automatic,scells2021comparison,scells2020automatic,scells2020sampling}.

In evaluating the variability of effectiveness across different generation models, we observe a difference in terms of different topic types. In the case of DTA topics, the Alpaca model shows a higher degree of variability compared to the ChatGPT model. This is evidenced by a higher variance observed in the CLEF-2017, CLEF-2018, and CLEF-2019-DTA datasets, where the variance of the Alpaca model is 14.3\%, 7.7\%, and 166\% greater than that of the ChatGPT model, respectively. 
On the other hand, for intervention topics, or topics in the Seed Collection that are not classified, the Alpaca model demonstrates more stability. Specifically, its variance is 39.1\%, and 28.6\% lower than that of ChatGPT.

Upon examining the average effectiveness, the fusion of multiple generations generally outperforms Single-Generation. Exceptions occur in the CLEF-2017 and CLEF-2019-DTA datasets with ChatGPT queries, and in the CLEF-2019-Intervention dataset with Alpaca queries. Moreover, the fusion of Multi-Generations from the Alpaca model consistently performs better than Boolean queries.

\enlargethispage{2\baselineskip}
Without a doubt, Multi-Generation Oracle queries consistently achieve the highest effectiveness, marking a considerable margin over the other ranking methods. This tells that with a proper technique or investigation to know how to select the best query over Multi-Generations, it could potentially lead to significant improvements in the effectiveness of screening prioritisation tasks.

\subsection{Ablation Studies}
\label{subsec:ablation}

To gain deeper insights into why generating a natural language query could yield higher effectiveness, and to understand the role of fusion, and the training process in the effectiveness of screening prioritisation, we conduct a series of ablation studies that investigates these factors.

\subsubsection{Generate query vs generate title}

\begin{table}
\centering
\small
\caption{Results comparing the effectiveness of generating a title (GT) versus generating a natural language query (GQ) from the Boolean query of a systematic review for screening prioritisation. Statistical significant differences ($p < 0.05$) between the effectiveness of a generated title versus a generated natural language query are indicated by~$*$.}
\label{table:query versus title}
\renewcommand{\tabcolsep}{6.9pt}
\renewcommand{\arraystretch}{0.95}
\begin{tabular}{@{}llclll@{}}
\toprule
\bf Dataset&\bf Model&\bf Query&\multicolumn{1}{@{}c@{}}{\bf AP}&\kern-0.4em\bf WSS95&\kern-0.7em\bf WSS100 \\ \midrule
\multirow{4}{*}{CLEF-2017}  & ChatGPT & GQ & \textbf{0.293} & \textbf{0.590} & \textbf{0.501} \\
& ChatGPT &  GT& 0.140* & 0.486* & 0.396* \\
\cmidrule(l@{\tabcolsep}){2-6}
& Alpaca & GQ & \textbf{0.284} & 0.591 & \textbf{0.502} \\
& Alpaca & GT & 0.270 & \textbf{0.595} & 0.502 \\

\midrule
\multirow{4}{*}{CLEF-2018} & ChatGPT & GQ & \textbf{0.381} & \textbf{0.692} & \textbf{0.528} \\
& ChatGPT & GT & 0.277* & 0.626* & 0.491 \\
\cmidrule(l@{\tabcolsep}){2-6}
& Alpaca & GQ & \textbf{0.333} &\textbf{0.640} & 0.485 \\
& Alpaca & GT & 0.307 & 0.637 & \textbf{0.501} \\

\midrule
\multirow{4}{*}{CLEF-2019-DTA} & ChatGPT & GQ & \textbf{0.247} & \textbf{0.660} & \textbf{0.528} \\
& ChatGPT & GT & 0.175 & 0.565* & 0.504 \\
\cmidrule(l@{\tabcolsep}){2-6}
& Alpaca & GQ & \textbf{0.241} & \textbf{0.665} & \textbf{0.521} \\
& Alpaca & GT & 0.164 & 0.544* & 0.458\\

\midrule
\multirow{4}{*}{\parbox{1.5cm}{\raggedright CLEF-2019-Intervention}} & ChatGPT & GQ & \textbf{0.433} & \textbf{0.573} & \textbf{0.503} \\
& ChatGPT & GT & 0.164* & 0.443* & 0.404 \\
\cmidrule(l@{\tabcolsep}){2-6}
& Alpaca & GQ & \textbf{0.317} & \textbf{0.491} & \textbf{0.448} \\
& Alpaca & GT & 0.232 & 0.458 & 0.408 \\

\midrule
\multirow{4}{*}{Seed Collection} & ChatGPT & GQ & \textbf{0.217} & \textbf{0.530} & \textbf{0.505}\\
& ChatGPT & GT & 0.127* & 0.494 & 0.490 \\
\cmidrule(l@{\tabcolsep}){2-6}
& Alpaca & GQ & \textbf{0.221} & \textbf{0.529} & \textbf{0.500} \\
& Alpaca & GT & 0.164 & 0.432* & 0.439 \\

\bottomrule
\end{tabular}
\end{table}

In our first ablation experiment, our underlying intuition for generating a natural language query instead of a systematic review title from the Boolean query is that we believe a title may only cover a narrow aspect of the Boolean query. Therefore, if vital information from the title is missed, it could result in lower effectiveness. To test this assumption, we compare generating a systematic review title and a natural language query from a Boolean query for the task of screening prioritisation.

To accomplish this, we first train a cross-encoder BioBERT model using the training portions of each dataset to rank documents using the final review title. For generating the review title, we employ ChatGPT in a zero-shot fashion, as fine-tuning the model is not yet available. For Alpaca, we fine-tune the model using the review titles in the training portion of our dataset using the same parameters as described in Section~\ref{sec:bool_query_conversion}, and then test on the testing portion.

Our results, presented in Table~\ref{table:query versus title}, clearly demonstrate that generating titles almost always yields lower effectiveness than generating natural language queries, regardless of whether the generation is done using ChatGPT or Alpaca (with the only exceptions being WSS95 on CLEF-2017 and WSS100 on CLEF-2018 when comparing the Alpaca model). Nevertheless, generating titles using ChatGPT appears to be significantly lower than when generated through the Alpaca model, with most results showing statistical significance.

\subsubsection{Impact of Fusion}

We further explore how the fusion of results from both Boolean and generated queries impacts the effectiveness of screening prioritisation. In Figure~\ref{fig:ablation_gain_loss}, we compare the effectiveness of using Boolean queries and generated queries separately versus using their fused results for screening prioritisation.

The results indicate that the fusion, on average, consistently outperforms using the generated query alone, but it is not always more effective than using the Boolean queries alone. The effectiveness of Boolean queries should not be overlooked. When comparing results across the two generation models, we observe that the effectiveness gains over Boolean queries obtained tend to be more stable when ChatGPT is used. Using ChatGPT in query generation may thus contribute to more consistent improvements when the results are combined with those from Boolean queries.

\subsubsection{Train Ranker using Single-Generation or Multi-Generations}

\begin{figure}
\includegraphics[width=\columnwidth]{fig/plot-ablation-gain-loss}
\caption{Differences in MAP from Boolean, Generated Query (GQ) to their fused effectiveness. }
\label{fig:ablation_gain_loss}
\vskip7.5ex
\includegraphics[width=\columnwidth]{fig/plot-combined-ablation}
\caption{Effectiveness when different training and inference settings are used for ranking candidate documents using the generated natural language query from ChatGPT.}
\label{fig:ablation_train}
\end{figure}

In our experiments, we train our natural language query-based ranker using Single-Generation results from generation models for reproducibility purposes, as Multi-Generations setup do not yield deterministic results each time, even when given the same prompt. However, we are interested in understanding how using Single-Generation versus Multi-Generations impacts the final outcome of the trained ranking model. To explore this, we formulated four distinct training and inference strategies for our downstream ranking model, which we refer to as Single-Train, Multi-Train, Single-Inference, and Multi-Inference.

For Single-Train, we train our model using the Single-Generation result from each Boolean query. For Multi-Train, we incorporate all generations from the generation model for each Boolean query in our training data. For Single-Inference, we test our model using a Single-Generation result from each Boolean query. Lastly, for Multi-Inference, we test our model using all generated queries from each Boolean query, and fuse them together.

With the same training parameters applied, we present the resulting effectiveness of screening prioritisation from ChatGPT using a bar chart in Figure~\ref{fig:ablation_train}. From the results, it is apparent that training the neural ranker using multiple creative queries does not typically yield higher effectiveness compared to training on a single deterministic query. The sole exception to this observation is the CLEF-2019-DTA dataset. However, when it comes to Multi-Inference, models generally exhibit improved effectiveness. This implies that the diversity introduced in the inference stage can positively impact the effectiveness of the ranking model, allowing it to generalise better and handle different query formulations. On the other hand, training using a diverse number of generated queries for the same topic may not significantly improve the effectiveness of the ranking model. This is likely due to the model being trained to generalise over multiple query formulations, which could lead to an averaging effect on the learned query-document relevance patterns.

\section{Summary of Findings}

Finally, we answer our research questions based on our results:

\paragraph{{\bf RQ1:} Comparison of original Boolean queries and generated natural language queries.}
We find that generating natural language queries generally results in higher effectiveness than using Boolean queries. This is valid both when the Boolean query is used in the context of the SOTA neural rankers for screening prioritisation, and when used within the previously proposed CLF technique~\cite{scells2020clf}, the only published technique for screening prioritisation that explicitly uses the Boolean query for ranking.

We also find that large gains can be obtained when the rankings obtained when using the original Boolean query and the generated natural language query are fused together. This result was obtained when using a simple rank fusion method, CombSUM: further improvements might be possible if using more sophisticated fusion methods~\cite{zhou2010information, wang2021bert}. This result suggests that these queries have complementary characteristics that can benefit screening prioritisation.

\paragraph{{\bf RQ2:} Impact of generation models.}
When the effectiveness of two generation models, ChatGPT and Alpaca, are compared, we observe that ChatGPT consistently generates natural language queries that are more effective in screening prioritisation. The gap in effectiveness is more pronounced for the CLEF-2019-Intervention dataset. This may be attributed to the training of Alpaca models on primary DTA topics, with intervention topics only contained in the test portion. This could have affected Alpaca's ability to perform effectively on the intervention topics. Conversely, ChatGPT, used in a zero-shot fashion, is not specifically tailored towards any topic; thus, its effectiveness is not significantly influenced by different types of systematic reviews.

\paragraph{{\bf RQ3:} Impact of ranking methods.}
We find that neural methods consistently outperform traditional term-matching methods; they also outperform runs submitted by the research teams that participated in the CLEF TAR shared tasks associated with the datasets we use. This finding highlights the robustness and effectiveness of neural ranking methodologies for the screening prioritisation task.

\paragraph{{\bf RQ4:} Effect of Multi-Generations.}
We identify considerable variance in the effectiveness of multiple natural language queries generated from both ChatGPT and Alpaca when applied to screening prioritisation. Notably, the Alpaca model tends to generate more unstable queries. However, when the results derived from these diversified queries are integrated, they often outperform the strategy of generating and using just one deterministic query for ranking documents. This occurs in~52.3\% of cases when using ChatGPT and in~61.7\% of cases when using Alpaca. If we also consider instances where the effectiveness is tied, these percentages increase to~55.5\% for ChatGPT and~70.3\% for Alpaca.

This finding suggests that the creativity of generative LLMs can enhance the natural language query generation task. Importantly, our findings also indicate that if a method could be implemented to effectively select the highest-performing generated query, the effectiveness of the downstream screening prioritisation task can be significantly improved (Oracle results). This potential for query selection may open new avenues for improving systematic review processes, pointing to the value of research into query performance predictors for systematic reviews; research on query performance predictors has been substantial in general information retrieval~\cite{arabzadeh2021bert}, but very scarce in the context of systematic reviews, where common predictors have been shown to be mostly ineffective~\cite{scells2018query}.

\paragraph{{\bf RQ5:} Derived natural language queries vs.\ working titles.}
We find that using a systematic review's working title as an input query for screening prioritisation generally results in lower effectiveness when compared to the use of our methods that uses the Boolean query for the same review to derive a natural language query to rank the candidate documents. This is different from when the final titles of the review are used: a practice that is common when experimenting with automation methods for systematic review, but only possible in retrospective evaluation, and not in practise. This discrepancy between working title effectiveness and final title effectiveness may be due to the evolving nature of the review title throughout the research process for the systematic review. 

\section{Conclusion}

Our approach to screening prioritisation advances the state of the art by combining the power of large language models, neural rankers, and relying only on information available at the time of screening during the production of a systematic review. Previous work relied instead on the final review title as the query for ranking candidate documents for screening, which is only available at the end of producing a systematic review. This led to overestimated effectiveness scores, as our experiments show. Using instruction-based LLMs to generate queries from the Boolean queries available at the time of screening is competitive with the state of the art using the final title. We also show that improvements in effectiveness can be achieved when rankings based on Boolean queries and generated natural language queries are combined with rank fusion.

Our results also show that while Alpaca, an open-source generation model, can match ChatGPT's effectiveness in some cases, ChatGPT generally produces better natural language queries, leading to more effective screening prioritisation. We also found that multiple generations of natural language queries, while leading to high variance in effectiveness, have the potential to yield a significant increase in effectiveness when effective query performance predictors are available to identify the best query variants, which leaves room for future work.

In summary, this paper has demonstrated the value of instruction-based models in generating and improving queries for screening prioritisation with neural rankers. Our future work involves investigating the potential of combining the query generation capability of instruction-based models with the highly effective ranking capability of neural rankers. In short, we believe that end-to-end training of instruction and ranking models can lead to even higher effectiveness in ranking documents.

\begin{acks}

Shuai Wang is supported by a UQ Earmarked PhD Scholarship. This research is funded by the Australian Research Council Discovery Projects programme ARC DP 210104043, and by the Universities Australia – DAAD Joint Research Co-operation Scheme. This work was partially funded by the European Commission under GA 101070014 (OpenWebSearch.EU).

\end{acks}
\balance
\bibliographystyle{ACM-Reference-Format}
\interlinepenalty=10000
\bibliography{sigir-23-generating-queries-chatgpt1.bib, sigir-23-generating-queries-chatgpt2.bib}

\end{document}